\documentclass[twocolumn, switch]{article} 

\usepackage{preprint}

\usepackage{amsmath, amsthm, amssymb, amsfonts}

\usepackage[numbers,square]{natbib}
\usepackage[utf8]{inputenc}	
\usepackage[T1]{fontenc}	
\usepackage{xcolor}		
\usepackage{booktabs} 		
\usepackage{nicefrac}		
\usepackage{microtype}		
\usepackage{lineno}		
\usepackage{float}			

\usepackage{lipsum}		

\usepackage{newfloat}
\DeclareFloatingEnvironment[name={Supplementary Figure}]{suppfigure}
\usepackage{sidecap}
\sidecaptionvpos{figure}{c}

\usepackage{titlesec}
\titlespacing\section{0pt}{12pt plus 3pt minus 3pt}{1pt plus 1pt minus 1pt}
\titlespacing\subsection{0pt}{10pt plus 3pt minus 3pt}{1pt plus 1pt minus 1pt}
\titlespacing\subsubsection{0pt}{8pt plus 3pt minus 3pt}{1pt plus 1pt minus 1pt}

\title{A Soft Recommender System for Social Networks}

\usepackage{xwatermark}
\newwatermark[firstpage,color=gray!60,angle=90,scale=0.32, xpos=-4.05in,ypos=0]{\href{https://doi.org/}{\color{gray}{Publication doi}}}
\newwatermark[firstpage,color=gray!60,angle=90,scale=0.32, xpos=3.9in,ypos=0]{\href{https://doi.org/}{\color{gray}{Preprint doi}}}
\newwatermark[firstpage,color=gray!90,angle=0,scale=0.28, xpos=0in,ypos=-5in]{*correspondence: \texttt{azrabiee@gmail.com}}

\usepackage{authblk}

\author[ ]{Marzieh Pourhojjati-Sabet}
\author[ \thanks{\tt{azrabiee@gmail.com}}]{Azam Rabiee}
\affil[ ]{Department of Computer Science, Dolatabad Branch, Islamic Azad University, Isfahan, Iran}

\usepackage{pifont}
\newcommand{\cmark}{\ding{51}}%
\newcommand{\xmark}{\ding{55}}%
\usepackage{makecell}

\begin{document}

\twocolumn[ 
  \begin{@twocolumnfalse} 
  
\maketitle

\begin{abstract}
Recent social recommender systems benefit from friendship graph to make an accurate recommendation, believing that friends in a social network have exactly the same interests and preferences. 
Some studies have benefited from hard clustering algorithms (such as K-means) to determine the similarity between users and consequently to define degree of friendships. 
In this paper, we went a step further to identify true friends for making even more realistic recommendations. we calculated the similarity between users, as well as the dependency between a user and an item. 
Our hypothesis is that due to the uncertainties in user preferences, the fuzzy clustering, instead of the classical hard clustering, is beneficial in accurate recommendations. 
We incorporated the C-means algorithm to get different membership degrees of soft users' clusters. 
Then, the users' similarity metric is defined according to the soft clusters. Later, in a training scheme we determined the latent representations of users and items, extracting from the huge and sparse user-item-tag matrix using matrix factorization.
In the parameter tuning, we found the optimum coefficients for the influence of our soft social regularization and the user-item dependency terms.    
Our experimental results convinced that the proposed fuzzy similarity metric improves the recommendations in real data compared to the baseline social recommender system with the hard clustering. 

\end{abstract}
\vspace{0.35cm}

  \end{@twocolumnfalse} 
] 



\section{Introduction}
\label{section:intro}
When you travel or shop, you listen, intentionally or unintentionally, to either the advice of experts in the field, or the recommendation of a friend who poses similar interests or tastes with you. 
Recently, obtaining advice or recommendation in social networks is getting popular and frequent. 
The widespread use of social networks has led to a faster pace of production of new information and resources in cyberspace, which itself confuses users to find or select the information and resources they need. 
The recommender system (RS) as a useful tool can offer specific and useful information from a large volume of information to suit the user's interest and taste.

The abundance and popularity of social networks, as well as the unique acceptance of users on social networks have made social interactions among users a more accurate interpretation of their preferences.
Researchers regard the social interactions as a useful source of information to improve the quality of their recommendations. As a result, a variety of \textit{social-network-based recommender systems} have been proposed in recent years. In the rest of the paper, we name it \textit{social recommender system (SRS)} for the sake of simplicity.

\begin{figure}[H]
  \centering
  \includegraphics[width=\linewidth]{"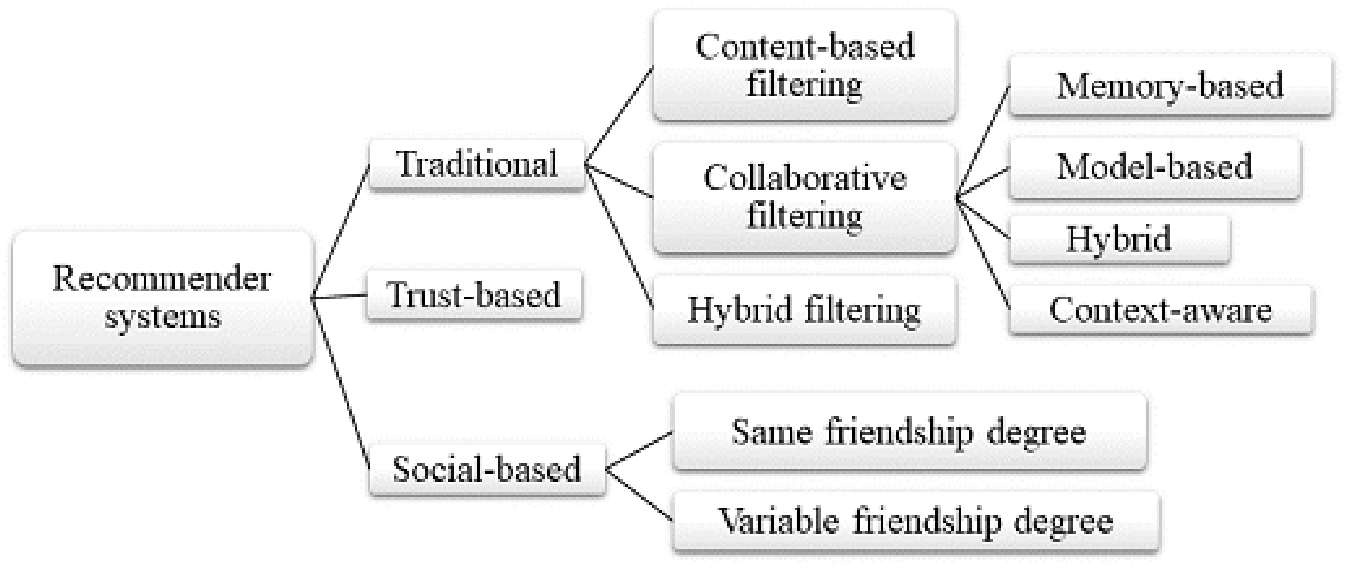"}
  \caption{Taxonomy of recommender systems.}
  \label{fig:taxonomy}
\end{figure}

Before the advent of social networks, traditional recommender systems focused on the user-item-rating (or user-item-tag depending on the application) matrix without considering the users friendship, as the information were not available.
Figure \ref{fig:taxonomy} depicts a taxonomy of recommender systems in which the first category is referred to as the traditional RS. 
Well-known techniques of this category are content-based \cite{van2000using, lops2011content} and collaborative filtering \cite{ekstrand2011collaborative, schafer2007collaborative, bergner2012model}.

Content-based approaches use only tags (ratings) of the same user to make a recommendation for her/him. These simple approaches are useless for a new user without previous ratings that is known as \textit{the cold-start problem}.

Collaborative filtering methods, on the other hand, use user-rating information either by memory-based (similar to the K-nearest neighbor method) \cite{zarei2019memory} or model-based  algorithms \cite{su2009survey}. 
A large group of research considered hybrid filtering by combining the two former filtering approaches \cite{gong2009combining}. 
Knowledge-based \cite{trewin2000knowledge} and demographic \cite{pazzani1999framework} filtering, also known as \textit{context-aware} collaborative filtering fall in this category, too.

The primary collaborative filtering methods usually suffered from the sparsity of the user-rating information. 
However, recent model-based collaborative filtering methods have focused on machine learning and deep learning architectures such as autoencoders to extract the low-dimension latent features from the sparse information \cite{sedhain2015autorec}.

Shown in Figure \ref{fig:taxonomy} as the second category, a few trust-based RSs \cite{massa2007trust, massa2004trust, o2005trust, golbeck2008computing} have been proposed, benefiting from users trusts. The systems mostly relied on the unilateral trust relation; while the social relation is cooperative and mutual. 
That weakness together with the limited access to the trust-aware datasets made trust-based systems to be impractical.

With the development of Web 2.0 technology and the abundant of datasets including the bilateral friendship information, many social-based RSs have been proposed \cite{sun2019leveraging}. 
As the third category shown in Figure \ref{fig:taxonomy}, many SRSs consider the social relationship among users assuming that the user and her friend have similar interests and tastes. 
Former SRSs assumed that friends are correlated anyway \cite{wang2014research}. 
So, they considered the same friendship degree for all the users in a group of friends; but users live in groups of people who are widely distributed, such as family, neighbors, classmates, and so on. 
Some of these people have interests in common, and others may even disagree her/him.

In fact, primary social RSs incorporate interests and ratings (tags) of the group of friends to determine the recommended item for users in the group; while typically friends may have different or even conflicting interests; or even if they focus on the same thing, their degree of interests may vary. 
Therefore, the new trend of SRSs has emerged considering variable friendship degree \cite{sun2015recommender, ma2011recommender, xu2018novel}. 

In this paper, we presented a soft (fuzzy) recommender system  that falls in this category, meaning that we defined a variable friendship degree and the corresponding user similarity measure to make more accurate and realistic recommendations. 
To summarize, our contribution is in two folds: (1) we introduced a soft users' similarity metric to have a realistic friendship degree for making the better recommendation, and (2) we defined a new user-item dependency term such that imposing it in a training scheme leads to better user and item latent representation. Details are explained in following sections.   

The rest of the paper is organized as follows. In Section \ref{sec:social-RS}, we  review a typical SRS that is our basedline algorithm. 
Later, Section \ref{sec:proposed-RS} explains our modifications and suggestions to make a soft recommendation by replacing the hard clustering (and consequently the user similarity metric) in the baseline algorithm.
Section \ref{sec:experimens} describes our experimental setup, explaining the dataset and evaluation metrics. Then, results come in Section \ref{sec:results}. We analyze impacts of parameters, and compare the performance of the baseline and the proposed soft recommendation in Section \ref{sec:results}. Eventually, conclusion comes in Section \ref{sec:conclusion}.

\section{Social Recommender System}
\label{sec:social-RS}

In this section, we explain our baseline SRS relying on collaborative filtering with a social regularization technique \cite{sun2015recommender, ma2011recommender, xu2018novel}. Generally, SRSs utilize the user-item-tag information together with the users friendship graph. It is worth mentioning that the method is applicable when rating (instead of tagging) is available, too. Without loss of generality, we consider that the user-item-tag matrix, denoted by $T$, together with the user friendship graph $U$ is available.  
In fact, the system is mostly relied on the collaborative filtering technique on the user-item-tag side, as well as a social regularization on the friendship graph side \cite{ma2011recommender}. 

In both collaborative filtering and social-based systems, dealing with large datasets and the sparse $T$, is not efficient. Thus, low-dimensional matrix factorization methods have been proposed \cite{sun2015recommender, xu2018novel, xu2019similarmf, koren2009matrix} to focus on $T$ matrix decomposition into low-dimensional user and item latent feature matrices, denoted by $S$ and $V$, respectively as follows, 
\begin{equation}
T \approx S^TV, 
\end{equation}
where $T\in R^{p\times q}$, in which $p$ (number of users) and $q$ (number of items) are quite large values. 
However, the estimated low-dimensional latent features $S\in R^{l\times p}$ and $V\in R^{l\times q}$ ($l\ll min(p,q)$) are very efficient for making the recommendation. Indeed, it is assumed that only a small number of factors influences preferences, and with a proper training scheme the preferences can be shown up in the latent feature matrices.

The matrix factorization method estimates the latent features using the following objective function in a training scheme that we will explain soon.

\begin{equation}
\label{eq:obj_fnc1}
\frac{1}{2} \Vert T-S^TV \Vert^{2}_{F} 
\end{equation}

Reforming Equation \ref{eq:obj_fnc1} by considering $\Vert.\Vert^{2}_{F}$ as the Frobenius norm, the objective function is as follows,

\begin{equation}
\min_{S,V}\frac{1}{2} \sum_{u=1}^{p} \sum_{i=1}^{q} a_{ui}(T_{ui}-S_u^T V_i)^2,
\end{equation}
  
in which, $a_{ui}=1$ if user $u$ selects the item $i$; otherwise $a_{ui}=0$. Moreover, $T_{ui}$ reflects tags on the item $i$ labeled by the user $u$. 

To avoid overfitting and to impose sparsity, L2-norm of the latent features are added to the objective function as follows,

\begin{equation}
\label{eq:basic-objective}
\min_{S,V}\frac{1}{2} \sum_{u=1}^{p} \sum_{i=1}^{q} a_{ui}(T_{ui}-S_u^T V_i)^2 + 
\frac{\lambda_1}{2}\Vert S \Vert^{2}_{F} + \frac{\lambda_2}{2}\Vert V \Vert^{2}_{F},
\end{equation}
    
where $\lambda_1$ and $\lambda_2$ are positive constants. As we have mentioned earlier, minimizing the above mentioned equation with respect to $S$ and $V$ leads to low-dimensional vector representations for users and items such that they are more efficient and less-computationally complex for the recommender system. Yet, our baseline recommender system includes the similarity between users and usre-item dependency terms described in the following subsections.   

\subsection{The Similarity Between Users}
\label{sec:users-similarity}

We add the social regularization terms to the objective function (Equation \ref{eq:basic-objective}) regarding the users' friendships, similar to \cite{ma2011recommender}. In our model, users have different degree of friendship regarding their preferences. Hence, we investigate friends with different interests and calculate the similarity among users based on their interests. The users' similarity measure is defined for each pair of users such that the social regularization term is defined as follows,
  
\begin{equation}
\label{eq:social_reg}
\frac{\beta}{2}\sum_{u=1}^p\sum_{f \in F(u)} sim(u,f)\Vert S_u-S_f \Vert^{2}_{F},
\end{equation}

where $\beta$ is a positive constant reflecting the importance of the social regularization term. However, $F(u)$ denotes the friend set of the user $u$. 
Furthermore, $sim(u,f)$ shows the similarity between user $u$ and $f$ that is going to be explained more in the next subsection. 
Adding Equation \ref{eq:social_reg} to the objetive function ensures that when $sim(u, f)$ is large (meaning that $u$ and $f$ are very similar), then latent features in $S_u$ and $S_f$ are extracted such that to be as close as possible. 

The similarity metric $sim(u,f)$ as a weight parameter controls the influence of $\Vert S_u-S_f \Vert^{2}_{F}$, which is the difference of the low-dimensional representation for the desired users. 
In our baseline recommender system, we define $sim(u,f)$ similar to the study performed by Sun and his colleague \cite{sun2015recommender} as follows,

\begin{equation}
\label{eq:hard-sim}
       sim(u,f)=k\frac{1}{q}\sum_{i=1}^q cos(T_{ui}, T_{fi}),
\end{equation}
\[
k=
\begin{cases}
       \lambda  & \text{if $u$ and $f$ are in the same cluster}   \\
       1-\lambda   & \text{otherwise}
\end{cases}
\]

where, $0<\lambda<1$ and $cos(T_{ui}, T_{fi})$ denotes the cosine similarity between tag vectors of users $u$ and $f$ to the same item $i$. Obviously, $n$ is the total number of items that both users are selected or labeled.

In our baseline recommender system, we cluster users utilizing K-means algorithm. 
With the above mentioned similarity metric (Equation \ref{eq:hard-sim}), we determine $sim(u,f)$ as a linear combination of the similarity of the same-clustered and the different-clustered users. 
The constant coefficient of the linear combination, $\lambda$, is a number between zero and one. 
The more $\lambda$ is close to one, the more the in-cluster similarity affects the model.
In the implementation of the paper (similar to \cite{sun2015recommender}), we supposed $\lambda=0.8$ to highlight the influence of the same-clustered users.     
Please note that the similarity metric guarantees that friends are not necessarily similar. 
They may be friend with different or even conflicting preferences unless they are in the same cluster.

\subsection{The User-Item Correlation}
\label{user-item-corr}

\begin{figure}[H]
  \centering
  \includegraphics[width=0.5\linewidth]{"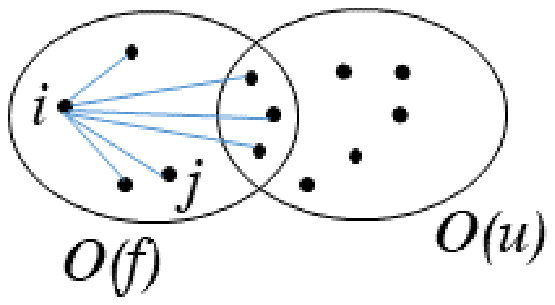"}
  \caption{The user $u$ and the item $i$ are correlated through the friend $f$.}
  \label{fig:user-item}
\end{figure}

In addition to the similarity between pairs of users, we consider the dependency between users and items, as well. 
the User-item dependency points out that how much the user $u$ relates to the item $i$. 
Xu \cite{xu2018novel} has considered the dependency only if the user $u$ has selected or labeled the item $i$; but the user and the item may be correlated even if they do not have any relation. 
To elaborate more, Figure \ref{fig:user-item} illustrates the case that the user $u$ is correlated to the item $i$ through the friend $f$. In the figure, $O(u)$ and $O(f)$ indicate sets of items selected by users $u$ and $f$, respectively.   

We introduce the user-item dependency similar to \cite{sun2015recommender}, in which the correlation is regardless of whether $i$ is selected by $u$ or not.
Sun and his colleague \cite{sun2015recommender} determined the dependency by two factors: (1) the similarity of the user $u$ and the friend $f$, as well as (2) the correlation of the item $i$ with other items selected by $f$. 
They defined the correlation with the latent representation of items ($V$); while mistakenly the derivative with respect to the $V$ has neglected it. 

Unlike \cite{sun2015recommender}, we define the user-item correlation with the user-item-tag matrix. 
In our definition, the user-item dependency is the distance of the user $u$ and the friend $f$ (denoted by $\Vert S_u-S_f \Vert^{2}_{F}$), times the average cosine similarity of the item $i$ and items $j$ (indicated by $corr(f, i)$).
Accordingly, Equation \ref{eq:user-item} introduces the user-item regularization term. 
  
\begin{equation}
\label{eq:user-item}
\begin{split}
\frac{\alpha}{2}\sum_{u=1}^p\sum_{i=1}^q\sum_{f \in F(u)} corr(f, i)\Vert S_u-S_f \Vert^{2}_{F},\\
corr(f,i)=\frac{1}{n}\sum_{j\in O(f)} cos(T_{fi}, T_{fj})
\end{split}
\end{equation}

where $\alpha>0$, $F(u)$ denotes the user $u$'s friends, and $corr(f, i)$ reflects the correlation of the friend $f$ and the item $i$ that may or may not be chosen by the user $u$. However, $j$s are the items selected by $f$, denoted by $j\in O(f)$, and $n$ is the number of items ($n=\vert O(f)\vert$). 

Eventually, we impose both the users' similarity (Equation \ref{eq:social_reg}) and the user-item correlation (Equation \ref{eq:user-item}) to the basic objective function (Equation \ref{eq:basic-objective}). Later, in a training scheme, we minimize the function such that the optimum user and item low-dimensional representations are extracted. The next subsection explains the training algorithm.

\subsection{The training algorithm}
\label{sec:training}

In our baseline recommender system, we utilize the low-dimensional matrix factorization with social regularization terms to extract the optimum representation for users and items. As we have explained so far, the objective function is as follows,

\begin{equation}
\label{eq:final-objective}
\begin{split}
Loss(S,V)= & \frac{1}{2} \sum_{u=1}^{p} \sum_{i=1}^{q} a_{ui}(T_{ui}-S_u^T V_i)^2 + 
\frac{\lambda_1}{2}\Vert S \Vert^{2}_{F} + \frac{\lambda_2}{2}\Vert V \Vert^{2}_{F}\\
& +\frac{\alpha}{2}\sum_{u=1}^p\sum_{i=1}^q\sum_{f \in F(u)} corr(f, i)\Vert S_u-S_f \Vert^{2}_{F}\\
& +\frac{\beta}{2}\sum_{u=1}^p\sum_{f \in F(u)} sim(u,f)\Vert S_u-S_f \Vert^{2}_{F}.
\end{split}
\end{equation}

We apply gradient descent algorithm to the user latent features $S_u$ and the item latent features $V_i$ to get a local minimum of the objective function. The derivatives are as follows,

\begin{equation}
\label{eq:derivatives}
\begin{split}
\frac{\partial L}{\partial S_u} = & \sum_{i=1}^{q} a_{ui}(S_u^T V_i-T_{ui})V_i + 
\lambda_1 S_u \\
& + \alpha \sum_{i=1}^q\sum_{f \in F(u)} corr(f, i)(S_u-S_f) \\
& + \beta \sum_{f \in F(u)} sim(u,f) (S_u-S_f)\\
\frac{\partial L}{\partial V_i} = & \sum_{u=1}^{p} a_{ui}(S_u^T V_i-T_{ui})S_u + 
\lambda_2 V_i. \\
\end{split}
\end{equation}

Table \ref{tab:pseudo-code} presents the pseudo code of the training algorithm. 
Inputs are: the user-item-tag matrix ($T$), the learning rate ($\eta$) , the user-item correlation coefficient ($\alpha$), the users' similarity coefficient ($\beta$), maximum number of training iterations (max\textunderscore iter), as well as the constant coefficients to avoid overfitting ($\lambda_1$, $\lambda_2$ for user and item latent features, respectively).
 
\begin{table}[H]
 \caption{Pseudo code of the training algorithm}
  \centering
  \begin{tabular}{ll}
    \toprule
    \multicolumn{2}{l}{Input: $T$, $\eta$, $\alpha$, $\beta$, max\textunderscore iter, $\lambda_1$, $\lambda_2$} \\
    \multicolumn{2}{l}{Output: $S$, $V$} \\
    (1) &  Initialize $S$, $V$ with random small values\\
    (2) &  for max\textunderscore iter > 0\\
    (3) &  \hspace{3mm}for $T_{ui}$ in $T$ do\\
    (4) &  \hspace{6mm}calculate $\frac{\partial L}{\partial S_u}$ using Equation \ref{eq:derivatives}\\
    (5) &  \hspace{6mm} for $d=1,..., l$ do\\
    (6) &  \hspace{9mm} $S_{ud}\leftarrow S_{ud}-\eta \frac{\partial L}{\partial S_u}$\\
    (7) &  \hspace{6mm}end for\\
    (8) &  \hspace{6mm}calculate $\frac{\partial L}{\partial V_i}$ using Equation \ref{eq:derivatives}\\
    (9) &  \hspace{6mm} for $d=1,..., l$ do\\
    (10)& \hspace{9mm} $V_{id}\leftarrow V_{id}-\eta \frac{\partial L}{\partial V_i}$\\
    (11)& \hspace{6mm}end for\\
    (12)& \hspace{3mm}end for\\
    (13)& \hspace{3mm} max\textunderscore iter -{}-\\
    (14)& \hspace{3mm} break if the the objective function $L$ converges\\
    (15)& end loop \\
    \bottomrule
  \end{tabular}
  \label{tab:pseudo-code}
\end{table}

Outputs of the algorithm would be the optimum user and item latent features denoted by $S\in R^{l\times p}$ and $V\in R^{l\times q}$ matrices, respectively. 
As described in Table \ref{tab:pseudo-code}, in the first line, matrices are initialized by small values. 
Later, lines (2) to (15) describe the main loop of the training scheme that stops either after a specified number of iteration (max\textunderscore iter) or if the objective function $L$ converges (line 14). 

In the main loop, for every element $T_{ui}$ of $T$, lines (4) to (11) are repeated, meaning that the derivative $\frac{\partial L}{\partial S_u}$ is calculated so that $S_{ud}$ is updated according to the gradient descent algorithm (line 6). 
Similarly, we calculate $\frac{\partial L}{\partial V_i}$ for updating $V_{id}$ values in line (10).  

Minimizing the objective function using the gradient descent satisfies applied constraints, including the sparsity of $S$ and $V$, as well as the closeness of the latent representations of friends with similar preferences, which itself is interpreted by both the users' similarity and the user-item dependency.   

\section{Proposed Soft Recommender System}
\label{sec:soft-RS}

Considering variable friendship degree improved the accuracy of the recommendation \cite{sun2015recommender}. However, we believe that the similarity metric defined based on the hard clustering does not precisely reflect the true friendship.
Thus, in this study, we suggest the fuzzy (soft) clustering and consequently the fuzzy (soft) similarity metric. Accordingly, we named the proposed algorithm as the \textit{soft recommender system}.

In the classical clustering, each sample belongs to one and only one cluster and cannot be a member of two or more clusters.
In another word, clusters do not have overlap.
What if a user is similar to two or more clusters?
In reality, fuzzy clsutering reflects the true friendships as it is a natural representation of groups.

Our hypothesis is that generalizing the natural clustering into the baseline social recommender system would improve the recommendations.    
The baseline system uses basic K-means clustering to obtain the users' similarity. 
In followings, we show how replacing the fuzzy C-means algorithm improves the accuracy of the recommendations. 
In fuzzy clustering, each user can belong to more than one cluster with varying degree of membership.
The degree of membership is a value between zero and one.
Therefore, we introduce a new users' similarity metric regarding the fuzzy membership degrees in the next subsection. 

\subsection{Soft User Similarity Metric}
\label{sec:soft-sim}

The baseline recommender system (similarly references \cite{sun2015recommender, ma2011recommender, xu2018novel} benefits from the k-means clustering and consequently the similarity metric described in Equation \ref{eq:hard-sim} to measure variable friendship degrees. 
In the equation, the $\lambda$ parameter compromises between the similarity of friends who are in the same cluster and friends who are not in the same cluster. 
The closer the $\lambda$ gets to one, the more important the same-cluster friends' similarity is.    

Regarding the equation, previous studies tried to compensate the uncertainty in friends' preferences with the fine-tuning of the $\lambda$ parameter. 
In the presented algorithm, the C-means as a fuzzy clustering estimates the membership degree of each user in each cluster, handling the uncertainty and consequently leading to more precise approximation of users interest and preferences. 

In K-means clustering, every user is assigned to the closest cluster utilizing the Euclidean distance. 
Even if a user is close to two or more clusters, the closets eventually is selected. 
Correspondingly, in C-means, the same user is assigned to all the close clusters but with varying membership degree. 
Let's assume $\mu_{uc}$ denotes the membership degrees of the user $u$ in the cluster $c$, and $\mu_{fc}$ is the same for the friend.   
The average of the friends' membership degrees' differences on all clusters, indicated by $\frac{1}{C}\sum_{c=1}^C\vert\mu_{uc}-\mu_{fc}\vert$, is an appropriate criterion for calculating the users' similarity.
The higher the criterion is, the more dissimilar the friends are.
 
In the proposed system, we formulate the users' similarity as follows,

\begin{equation}
\label{eq:soft-sim}
\begin{split}
  simF(u,f)=&(1-\frac{1}{C}\sum_{c=1}^C\vert\mu_{uc}-\mu_{fc}\vert)\\
  &\times\frac{1}{q}\sum_{i=1}^q cos(T_{ui}, T_{fi}),
\end{split}
\end{equation}
 
where $C$ is the total number of clusters. Besides, $\mu_{uc}$ and $\mu_{fc}$ are the membership degrees of the user $u$ and the friend $f$ in the cluster $c$, respectively. However, $T_{ui}$ and $T_{fi}$ are the tags labeled by the user $u$ and the friend $f$ to the item $i$.     

The proposed training algorithm is the same as the pseudo code described in Table \ref{tab:pseudo-code}, replacing $sim$ with $simF$ in Equation \ref{eq:derivatives}.

\section{Experiments}
\label{sec:experimens}

To compare the proposed soft recommender system with the baseline, we conducted  experiments by evaluating \textit{precision} and \textit{recall} metrics on \verb+Del.icio.us+ dataset \citep{basile_topical_2015} , described in following subsections.

\subsection{Evaluation Metrics}
\label{sec:eval-metrics}

We evaluated \textit{precision} and \textit{recall} according to Equation \ref{eq:eval-met}, in which $R(u)$ indicates the predicted tags the user $u$ may label to the items. 
‌Obviously, $T(u)$ is the actual tags the user $u$ labels.
  
\begin{equation}
\label{eq:eval-met}
\begin{split}
  precision=&\frac{\vert R(u)\cap T(u)\vert}{\vert R(u)\vert}\\
  recall=&\frac{\vert R(u)\cap T(u)\vert}{\vert T(u)\vert},
\end{split}
\end{equation}

The precision reflects the possibility that the user $u$ is interested in recommended items. 
It equals the ratio of the number of common predicted and actual items which $u$ labeled to the entire predicted recommended items. 
The recall is the same ratio to the actual recommended items, reflecting the possibility that an item which $u$ labeled may be recommended.

Clearly, the precision decreases if the number of recommended items increases.
Thus, for a fair comparison, we analyzed the precision with varying number of recommended items from 1 to 5. 
In results section, the notation $P@i$ specifies the precision when $i$ items are recommended. 
Likewise, $R@5$ determines the recall when 5 items are recommended.

\subsection{Dataset}
\label{sec:dataset}

We selected the real dataset \verb+Del.icio.us+ \cite{basile_topical_2015} to evaluate the soft recommender system for social networks.  
The dataset is an \verb+xml+ file, containing nearly 40 million records of users, URL resources as items, and tags. 
Users can label one or more tags to an item (a website) optionally.
For example, the user named "NicoDruif" may give tags "design", "usability", "inspiration", and "interactiondesign" to a website specified by a URL as the item. 
Moreover, users in the same group are considered as friends.

The size of the dataset is about one gigabyte, imposing heavy computations. 
We normalized the dataset by pruning users with the low number of labeled items and users with no friends.
Eventually, we created the user-item-tag matrix $T$ on the remaining 5000 users to compare the baseline and the proposed algorithms. 

\section{Results}
\label{sec:results}

We conducted some experiments in two folds: (1) fine tuning, evaluating the impacts of some parameters, and (2) performance analysis by comparing recommender systems. details are in following subsections. In the reported results, RSboSN (stands for Recommender System based on Social Networks) and F-RSboSN (for Fuzzy-RSboSN) refer to the baseline and the proposed models.    

\subsection{Fine Tuning}
\label{sec:fine-tune}

We evaluated the impact of some parameters, including the users' similarity coefficient $\beta$, the user-item correlation coefficient $\alpha$, and the latent feature dimension $l$. 
As the learning rate ($\eta$) is important in the training convergence and less effective in the performance, in all experiments $\eta$ equals $0.5$. 
Likewise, $\lambda$, $\lambda_1$, and $\lambda_2$ are also $0.5$. 
Moreover, the number of clusters in both baseline and the soft models is 10. 
In this subsection, we report the precision average when one item is recommended to each user ($P@1$). 
Moreover, we examined both baseline and proposed models in evaluating the impacts of parameters.   
 
\subsubsection{Impact of $\beta$}

\begin{figure}[H]
  \centering
  \includegraphics[width=\linewidth]{"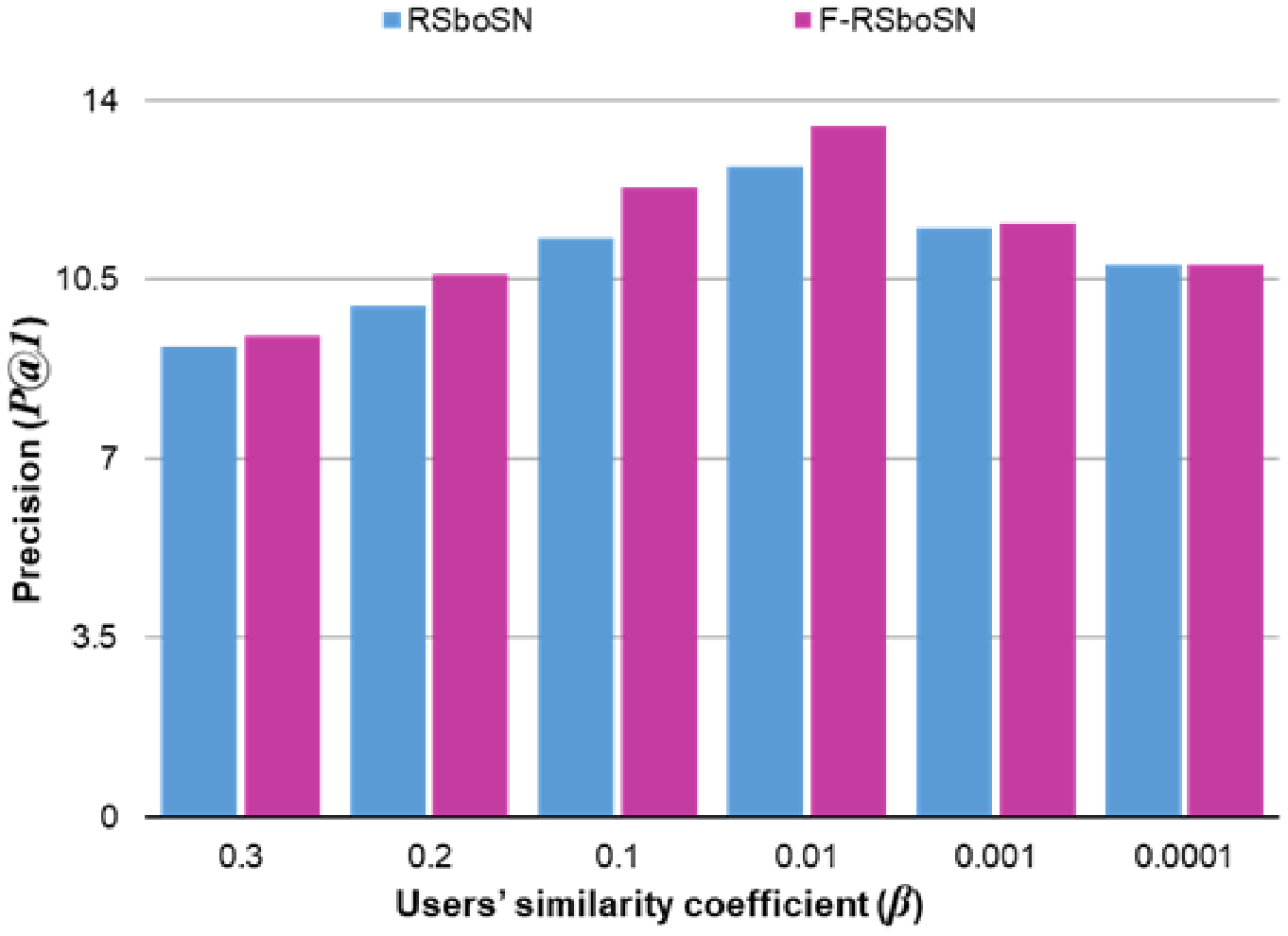"}
  \caption{The impact of the users' similarity coefficient $\beta$.}
  \label{fig:beta}
\end{figure}

Tuning an appropriate value for $\beta$ has a significant effect on the final performance. 
Therefore, we reported $P@1$ for both the baseline and the soft models when $\beta$ varies logarithmic from 0.0001 to 0.3 in Figure \ref{fig:beta}.   

As reported in Figure \ref{fig:beta}, the lower $\beta$ is, the closer the results of both models are. 
This is admissible as models are different in the users' similarity criterion, in which $\beta$ is its weight coefficient in the objective function.
Thus, decreasing $\beta$ diminishes the effect of the proposed users' preferences evaluation.
Meanwhile increasing the $\beta$ value to 0.1 and above suppresses other expressions in Equation \ref{eq:final-objective}, which itself reduces the performance.
According to Figure \ref{fig:beta}, an appropriate value for the parameter (0.01 and 0.1 in this case) leads to around 10\% improvement.     

\subsubsection{Impact of $\alpha$}

\begin{figure}[H]
  \centering
  \includegraphics[width=\linewidth]{"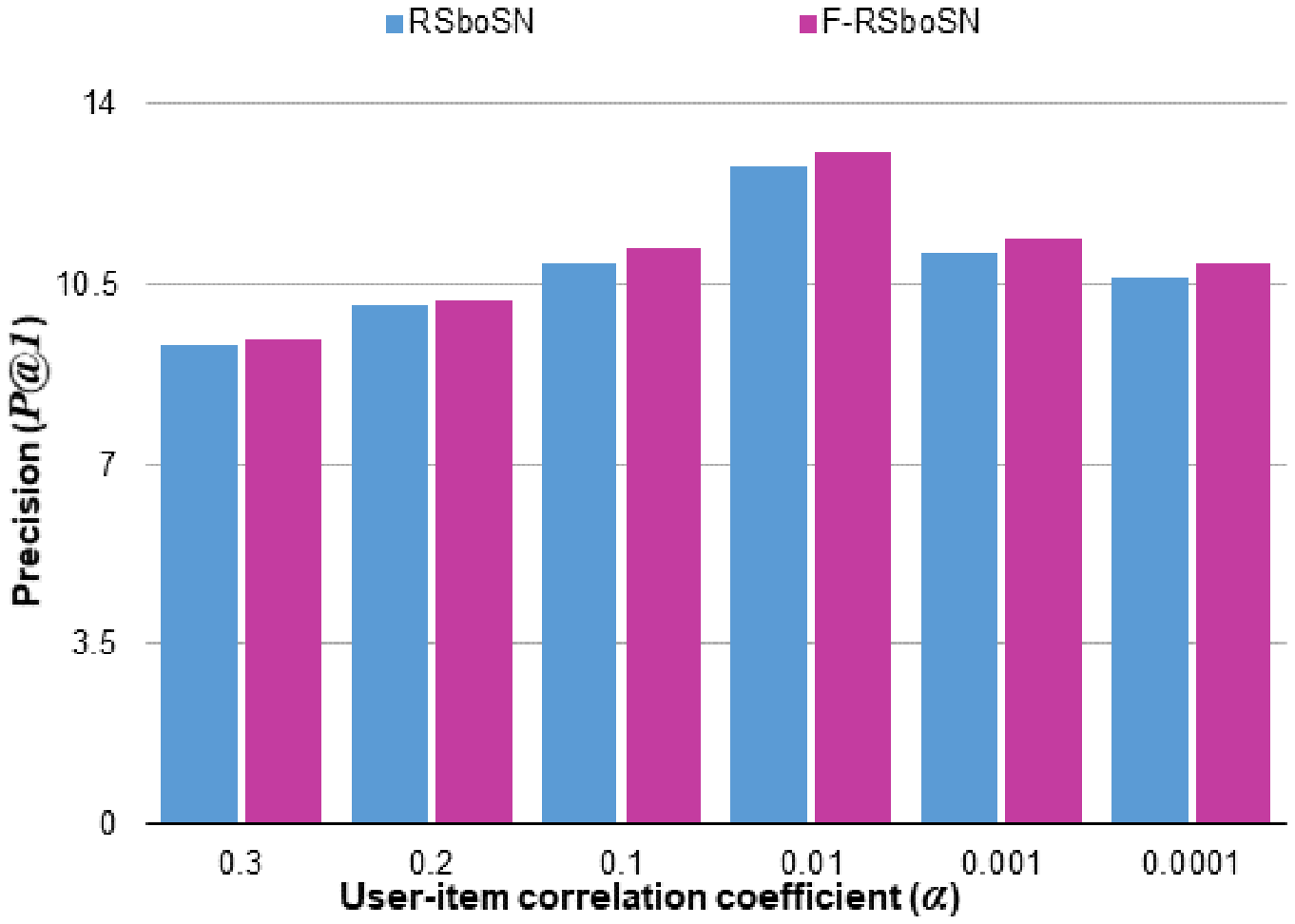"}
  \caption{The impact of the user-item correlation coefficient $\alpha$.}
  \label{fig:alpha}
\end{figure}

The user-item correlation term in the objective function (Equation \ref{eq:final-objective}) is the same for both the baseline and the proposed models. 
Thus, theoretically the precision of models should be the same. 
Our experiments also have convinced that with the same $\beta$ both models show the same performance. 
In the $\alpha$ analysis experiment, the $\beta$ value for RSboSN and F-RSboSN are 0.01 and 0.1, respectively. 
In addition, the latent feature dimension is set to 30.
Figure \ref{fig:alpha} illustrates the $P@1$ precision with respect to the coefficient $\alpha$ who varies logarithmic from 0.0001 to 0.3.     
As shown in the figure, changing $\alpha$ (regarding the fixed $\beta$ values) does not show significant differences for two models.
Yet, larger $\alpha$ (e.g. 0.2 and 0.3) dominates the user-item term in the loss function, suppressing the remaining terms that leads to worse performance. 
Similar to the previous experiment, empirically, we obtained 0.01 as the optimum value for $\alpha$.

\subsubsection{Impact of Feature Dimensionality}

We have already explained in Section \ref{sec:social-RS} that the low-dimensional latent representations for users and items are conductive for making the suggestions.    
The latent representations are products of the training algorithm from the sparse matrix $T$.
The latent condense features as the principle patterns are essential in finding new items to recommend.
Therefore, the latent feature dimension $l$ plays an important role in the performance. 
Thus, we evaluated the impact of the parameter in $P@1$ performance in another experiment.

\begin{figure}[H]
  \centering
  \includegraphics[width=\linewidth]{"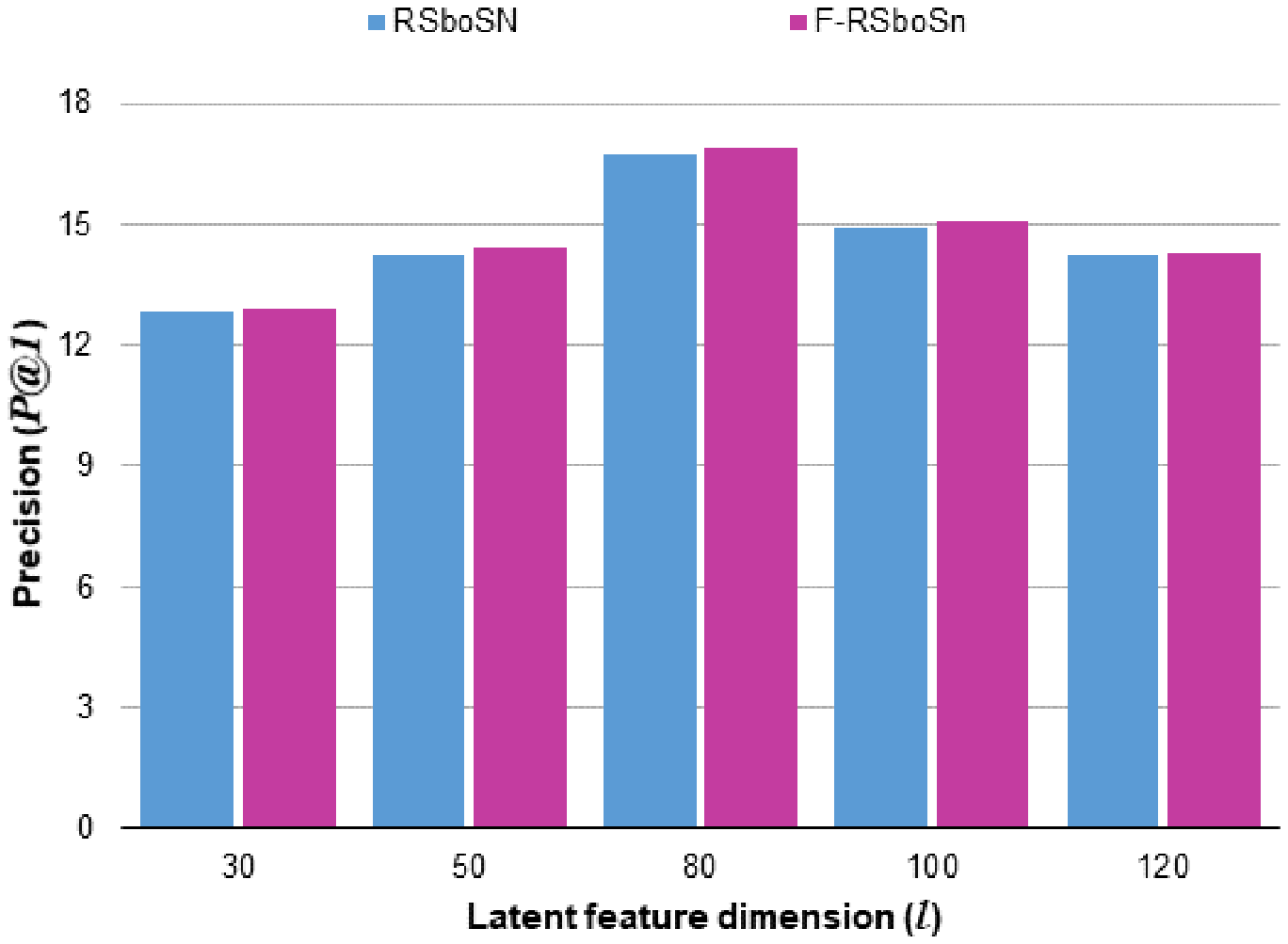"}
  \caption{The impact of the latent feature dimension $l$.}
  \label{fig:feature-dim}
\end{figure}

The large dimension implicates keeping more details or perhaps redundant information for every user and item, which is computationally complex.
On the other hand, small dimension may not be efficient, and may result in lower performance. Hence, our aim is to find the optimum dimension size for the latent representation by evaluating the values from 30 to 120. 
Figure \ref{fig:feature-dim} depicts the results of this experiment, convincing that with the fixed $\alpha$ and $\beta$, the models are almost the same. 
The $\alpha$ and $\beta$ are set to 0.01.
As the results has pointed out, the empirically-obtained optimum dimension size is 80.        

\subsection{Performance Analysis}
\label{sec:performance}

In order to show the performance improvement of the proposed approach, we compare the fuzzy recommender system with the following common methods: 
\begin{itemize}
\item the content-based filtering, also known as popularity approach (Pop),
\item the user-based collaborative filtering approach (u-CF), 
\item and the social regularization approach (SoReg) \cite{ma2011recommender}.
\end{itemize}
The content-based or popularity approach is to recommend the most popular items corresponding to the tags which the user $u$ labeled frequently. 
The method ignores users' friendships. Thus, the popularity of item $i$ for the user $u$ is simply defined as follows,

\begin{equation}
\label{eq:pop}
  p(i, u)=\sum_{t\in T(u)}n_{ut}n_{ti},
\end{equation}

in which, $T(u)$ is the set of tags that the user $u$ uses, $n_{ut}$ is the number of times the user $u$ used the tag $t$, and $n_{ti}$ is the number of times the item $i$ is labeled by the tag $t$.

The user-based collaborative filtering (u-CF) method defines the popularity of the item $i$ for the user $u$ as follows,
\begin{equation}
\label{eq:u-cf}
  p(i, u)=\sum_{v\in N(u)}a_{vi}sim(u, v),
\end{equation}
     
where $N(u)$ is the set of the neighbor users to the user $u$, and $sim(u, v)$ is the similarity of the user $u$ and his/her neighbor $v$ that is defined by a simple cosine similarity. Please note that the u-CF is not a social-based method; so the user friendship graph is not available; yet, users' similarity can be defined according to their tags. However, $a_{vi}=1$ if user $v$ selects the item $i$, otherwise $a_{vi}=0$. 

The social regularization (So-Reg) approach is very similar to our baseline except that the user-item dependency term is omitted in Equation \ref{eq:final-objective}. Therefore, the method is a social-based system with only the social regularization term such that its objective function is defined as follows,   

\begin{equation}
\label{eq:SoReg-objective}
\begin{split}
Loss(S,V)= & \frac{1}{2} \sum_{u=1}^{p} \sum_{i=1}^{q} a_{ui}(T_{ui}-S_u^T V_i)^2 + 
\frac{\lambda_1}{2}\Vert S \Vert^{2}_{F} + \frac{\lambda_2}{2}\Vert V \Vert^{2}_{F}\\
& +\frac{\beta}{2}\sum_{u=1}^p\sum_{f \in F(u)} sim(u,f)\Vert S_u-S_f \Vert^{2}_{F}.
\end{split}
\end{equation}

\begin{table}[H]
 \caption{Properties of recommender systems used in the performance comparison}
  \small  
  \centering
  \begin{tabular}{c|ccccc}
    \toprule
     approach 						& Pop 	& u-CF 	& SoReg & RSboSN & F-RSboSN \\
     \midrule     
     \makecell{users'\\ dependency} & \xmark & \cmark & \cmark & \cmark & \cmark \\   
     \midrule     
	 social-based 					& \xmark & \xmark & \cmark & \cmark & \cmark \\
     \midrule     
     \makecell{user-item\\ dependency}& \xmark &\xmark &\xmark & \cmark & \cmark \\   
     \midrule     
     \makecell{variable\\ friendship \\ degree}&\xmark &\xmark &\cmark & \cmark & \cmark \\   
     \midrule     
     \makecell{users' soft \\ grouping}&\xmark &\xmark &\xmark & \xmark & \cmark \\   
    \bottomrule
  \end{tabular}
  \label{tab:approaches}
\end{table}

\begin{figure}[H]
  \centering
  \includegraphics[width=\linewidth]{"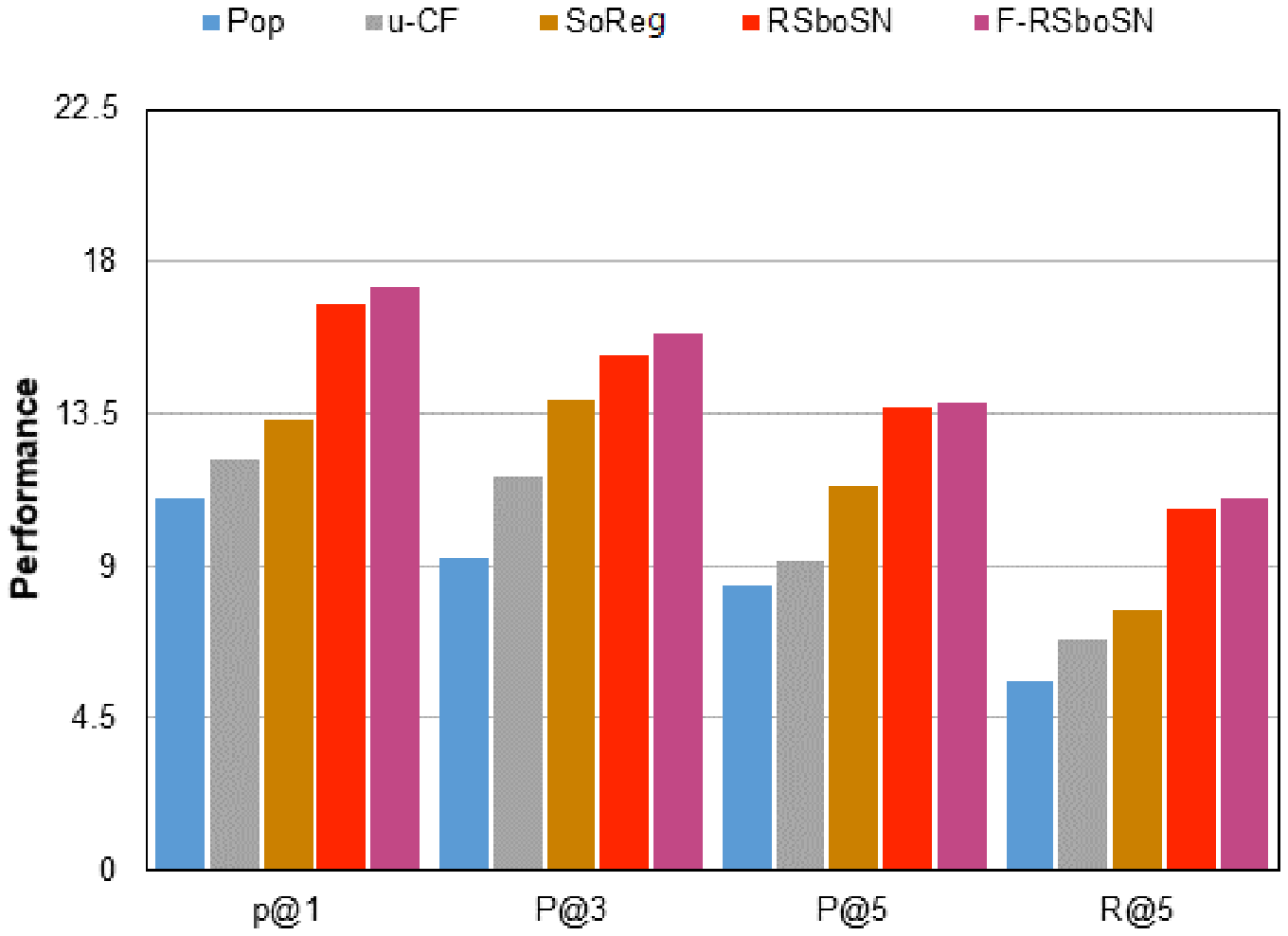"}
  \caption{Comparison of the performance of recommender systems.}
  \label{fig:comparison}
\end{figure}

Table \ref{tab:approaches} exhibits properties of the approaches used in the performance evaluation in one shot. 
For a comprehensive analysis, the opponent approaches are chosen from various categories of recommender systems.

Figure \ref{fig:comparison} compares the performance of the above mentioned approaches besides the baseline and the proposed models.  
We compared $P@1$, $P@3$, $P@5$, and $R@5$ that are precision when 1, 3, and 5 items are recommended, as well as the recall when 5 items are recommended.
In the performance comparison experiment, $l$, $\alpha$, and $\beta$ are 80, 0.01, and 0.01, respectively. 

For all four metrics, the proposed model is superior to the baseline and shows a significant improvement over the other methods. 
The content-based filtering (Pop) ignores any sort of users and user-item dependencies. 
Our experiments also explicitly confirmed its weak performance.    
Moreover, experiments approved that considering the user-item dependency is beneficial as both RSboSN and F-RSboSN models show supremacy over SoReg.

\section{Conclusion}
\label{sec:conclusion}

The uncertainty is users' friendship has motivated up to introduce a soft recommender system for social networks in which friends have different or even conflicing interests. 
The soft or fuzzy clustering presented in our approach shows more admissible approximation of users' similarity. 
Whereas, the previous similarity metrics have examined only two cases of the same-clustered and the different-clustered friends; then parameter-tuning for a $\lambda$ coefficient helped them to find the balance in the two cases. 
Our soft grouping have two notable features: (1) each user can belong to more than one cluster, and (2) the degree of membership in every cluster is a value between zero and one. 
According to our experiments, the two features were beneficial in making more appropriate recommendations.    

We examined the approach on a real dataset. 
The proposed model may behave differently in other datasets, depending on the number of users' friends. It seems that the larger the user friend group is, the more accurate the information is provided to the system. 
Thus, the proposed algorithm may work much more effectively because if the user friend group is restricted, the proposed algorithm will have less data to match.




\normalsize
\bibliography{references}


\end{document}